\documentclass[conference]{IEEEtran}
\usepackage{tabularx}
\usepackage{booktabs}
\usepackage{multirow}
\usepackage{doi}
\usepackage{hyperref}
\usepackage{cite}
\usepackage{amsmath,amssymb,amsfonts}
\usepackage{algorithmic}
\usepackage{graphicx}
\usepackage{textcomp}
\usepackage{xcolor}
\usepackage{pifont}% http://ctan.org/pkg/pifont
\usepackage{makecell}
\usepackage[flushleft]{threeparttable}

\newcommand{\cmark}{\ding{51}}%
\newcommand{\xmark}{\ding{55}}%
\def\BibTeX{{\rm B\kern-.05em{\sc i\kern-.025em b}\kern-.08em
    T\kern-.1667em\lower.7ex\hbox{E}\kern-.125emX}}
\begin{document}

\title{Financial Dynamics and Interconnected Risk of Liquid Restaking}

\author{
\IEEEauthorblockN{Hasret Ozan Sevim}
\IEEEauthorblockA{
University of Camerino (Camerino, Italy) \\
Catholic University of Sacred Heart (Milan, Italy) \\
hasretozan.sevim@unicam.it
}
\and
\IEEEauthorblockN{Christof Ferreira Torres}
\IEEEauthorblockA{
Instituto Superior Técnico (Lisbon, Portugal) \\
INESC-ID (Lisbon, Portugal) \\
christof.torres@tecnico.ulisboa.pt
}
}

\maketitle

\begin{abstract}
Decentralized finance introduces new business models and use cases as part of digital finance. Restaking has recently emerged as a transformative mechanism in DeFi, promising extra yields but introducing complex and interconnected risks. The paper monitors the current restaking landscape, empirically analyzes the revenue drivers of a liquid restaking protocol, and conducts a technical investigation on the emitted risk arising from the interconnection between liquid restaking and other protocols. The revenue dynamics of Renzo Protocol are analyzed by employing an OLS regression model, Granger-causality and random forest feature importance tests. Our results identify that revenue is primarily predicted by the value locked in the underlying EigenLayer ecosystem, the yield of Renzo protocol's liquid restaking token and the multi-blockchain expansion of that token. The multi-blockchain expansion of the liquid restaking token presents a double-edged sword: bridging to other networks is crucial for user adoption, but it adds the bridge risks to the existing risks of restaking. We investigate the cross-contamination risk between different DeFi services and the liquid restaking protocol. By mapping the asset flow across the decentralized finance ecosystem, it is detected that the bridge risk of the current size of Renzo's liquid-restaking assets does not impose a systemic risk on the current restaking and staking ecosystem. To address the potential consequences of the emphasized interconnection risks, we introduce two hypothetical scenarios and a stress test, assuming a large number of compromised liquid restaking tokens and a smart contract logic failure in a DeFi protocol. Considering the overall liquid-restaking protocols and the growing interconnection, this analysis requires further work to explore the growing complexities.
\end{abstract}


\section{Introduction}
The landscape of digital finance is undergoing a rapid transformation, continuously accelerated by a wave of open innovations. Within this fast-evolving landscape, innovation is constantly seeding new business models and expanding the use cases for different financial technologies. The primary example of this shift is the rise of decentralized finance (DeFi), which has emerged as the main financial application of blockchain and distributed ledger technologies (DLTs). By leveraging smart contracts, DeFi is involving new use-cases, services and business models on the DLTs from lending and borrowing to yield-bearing assets.

\begin{figure}[h]
    \centering
    \caption{Restaking phenomenon can be explained with 'multi-validation'. Validators also work for other networks or protocols, not only for Ethereum. So they give more service with the same operation. But restakers put their ETH assets at risk if the validation fails.}
    \includegraphics[width=0.48\textwidth]{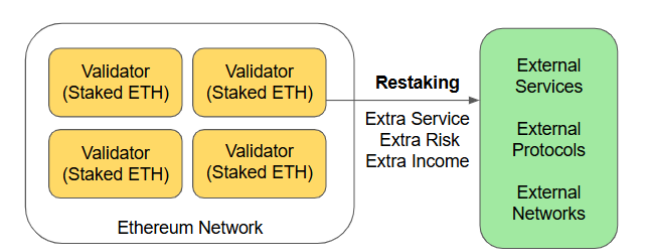}
    \label{fig:multivalidation}
\end{figure}

Since the Ethereum network has switched its consensus mechanism to proof-of-stake (PoS) from proof-of-work (PoW), we have witnessed innovative use cases such as staking and liquid staking \cite{b1}. Among these innovations, restaking has recently gained significant attention. The total value locked (TVL) in the restaking and liquid-restaking protocols is over 28 billion USD as of October 4, 2025 \cite{b23}. Restaking functions as a secondary service layer built directly on top of the foundational process of the core consensus mechanism for PoS blockchains, by enabling the reuse of already staked assets to secure additional applications and earn extra rewards. For example, on EigenLayer as a pioneer and leading restaking protocol, staked assets can be restaked to validate and secure other applications or services that benefit from Ethereum’s decentralized infrastructure. This validation service is called \textit{actively validated services} (AVSs) \cite{b2}. By contributing to an AVS, restakers seek an additional yield in exchange with risking the staked assets \cite{b3}. Liquid restaking protocols (LRPs) are platforms that allow users to earn compounding rewards by securing multiple blockchain services with their staked assets, while issuing liquid restaking tokens (LRTs) that represent this restaked position and its accrued yield, enabling the tokens to be used elsewhere in DeFi. The relationship between LRPs and LRTs is similar to the liquid staking model with liquid staking protocols (LSTs) and liquid staking tokens (LSTs). The paper aims to contribute insights to observe the restaking protocols, understand the revenue dynamics of liquid restaking, and the additional risk originating from bridging liquid restaking tokens as a part of a multi-blockchain revenue strategy.

The LRP market has experienced explosive growth, now representing tens of billions of dollars in value. However, its rapid expansion and technical complexity outpace a comprehensive understanding of its economic drivers and associated risks. This paper aims to fill this critical gap through a three-part analysis. First, we provide a catalog of the leading restaking protocols and their core attributes in a comparative table. Second, we analyze the revenue dynamics of a leading LRP (Renzo Protocol) that deploys its LRT (ezETH) across multiple blockchains. We employ an OLS regression model, Granger causality and random forest feature importance tests to identify the key factors influencing its revenue generation. Finally, we monitor the current status and the size of the bridge risk from transferring LRTs across chains and discuss how this risk may compound with other restaking risks like slashing. By combining the economic and security analyses, we can understand the full picture of liquid restaking since the business model reveals the powerful incentives for rapid multi-chain expansion, while the security assessment highlights the new systemic risks this growth creates. Ignoring one in favor of the other gives an incomplete view; the sustainability of the entire restaking market depends on a comprehensive outlook.

\section{Related Work}

Gogol et al. (2024)~\cite{b3} provide the systematized and necessary knowledge of staking, LSPs, LSTs, restaking, LRPs and LRTs. The study systematically compares the protocol mechanics, including node operator selection, staking reward distribution, and slashing. Empirical analysis of token performance reveals that protocol design and market dynamics impact token market value. Then the authors present the recent developments in restaking and discuss associated risks and security implications. Lastly, the authors review the emerging literature on liquid staking and restaking.~\cite{b3} provides the fundamental knowledge to investigate the deeper dynamics in the restaking field. To understand the dynamics of staking and liquid staking, we also benefit from~\cite{b4},~\cite{b5},~\cite{b6},~\cite{b7},~\cite{b8}, and~\cite{b9}.

Chaudhry et al. (2025)~\cite{b10} address an important gap by constructing and analyzing a market-implied LRT premium spread, benchmarked against a more mature LST for matched tenors. The study uses a primary source for yield data of the Pendle Finance protocol, from where the historical daily yield data is collected for a range of contracts, tenors, and chains during the period from January 2024 to July 2025.~\cite{b10} motivated us for the current paper for two main reasons. First of all, one of our goals is to investigate the multi-blockchain strategy of an LRP. Respectively,~\cite{b10} detects that the LRT of Ether.fi is bridged into and restaked on the Karak and Zircuit ecosystems. It is deployed across multiple chains including Ethereum Mainnet, Arbitrum, BNB Chain, and Berachain. But we focus on the Renzo Protocol~\cite{b11} rather than Ether.fi. Second, the author uses historical TVL data in the analysis by annotating a methodological discussion on TVL for studies on restaking and DeFi, similar to Luo et al. (2025)~\cite{b12}. The TVL data provides methodological challenges that stem from the inflated metrics due to the double-counting of interconnected protocols and bridging an LRT to another ecosystem. We adequately addressed this challenge and eliminated it by completing our analysis with the filtered TVL data. We also contribute a significant aspect by pointing out the LRT bridging risk, quantifying the interconnection between restaking and other DeFi services, and the potential interconnectedness between the risks to trigger each other. In academia, many risks of restaking are analyzed in various studies, especially~\cite{b13},~\cite{b14}, and~\cite{b15}. New mechanisms are proposed to mitigate the risks, like in~\cite{b16}. But we point out a different contagious risk specifically stemming from the bridged LRT tokens.

\section{Restaking Protocol Taxonomy}

Liquid restaking is the liquid staking equivalent for restaking. Users deposit their ETH, stETH, or other restakable assets into an LRP such as EtherFi, Kernel (Kelp) DAO, or Renzo Protocol. The protocol handles the technical complexity of restaking via EigenLayer and mints a liquid restaking token (e.g., eETH, rsETH, ezETH) to the user. This LRT is a yield-bearing liquid representation of a diversified basket of restaking positions, which can be further deployed across DeFi. A liquid restaking protocol generates revenue by earning fees on the extra rewards that users earn when their staked assets are used to secure additional services and applications beyond the base blockchain. Since an LRT is able to earn extra yield from the restaking AVS activity, LRTs generally have a premium price above the underlying staking and native assets. For instance, we observe that ezETH has a higher price than Lido's LST stETH and the native asset ETH.

\begin{table}[t]
\caption{Restaking Protocol Taxonomy}
\label{tab1}
\centering
\footnotesize
\begin{tabular}{l l c r c c}
\toprule
\textbf{Category} & \textbf{Protocol} & \textbf{LRT} & \textbf{TVL (\$)} & \textbf{Active On} & \textbf{Data*} \\
\midrule
\multirow{6}{*}{Infra.} 
 &  EigenLayer & & 19,00B & 1 chain(s) &  N.A. \\
 &  Babylon & & 7,07B & 1 chain(s) & N.A. \\
 &  Symbiotic & & 1,22B & 1 chain(s) & N.A. \\
 &  Karak & & 0,19B & 7 chain(s) & N.A. \\
 &  Jito & & 0,17B & 1 chain(s) & N.A. \\
 &  Solayer & & 0,09B &  1 chain(s) & N.A. \\
\midrule
\multirow{7}{*}{LRP}
 & Ether.Fi & eETH & 11,21B & 3 chain(s) & \xmark \\
 &  Kernel DAO & rsETH & 1,97B & 16 chain(s) & \xmark \\
 &  Renzo & ezETH & 1,55B & 12 chain(s) & \cmark \\
 &  Mantle & cmETH & 0,63B & 1 chain(s) & \xmark \\
 &  Mellow & N.A. & 0,44B & 4 chain(s) & \xmark \\
 &  Puffer & pufETH & 0,18B & 1 chain(s) & \xmark \\
 &  Swell & rswETH & 0,10B & 1 chain(s) & \xmark \\
\midrule
Yield & Pendle & & 0,04B & 2 chain(s) & N.A. \\
\bottomrule
\end{tabular}

\vspace{0.3em}
\raggedright
\scriptsize
This table shows the taxonomy of the restaking protocols as of October 4, 2025. The restaking service was mainly launched on Ethereum. But currently, there is restaking infrastructure in other layer-1 networks such as Solana and Bitcoin, and LRTs are deployed on layer-2 networks to access a wider user base. Source: DeFiLlama (Last Access: October 4, 2025). 

*The checkmark symbol (\cmark) indicates sufficient protocol data availability for multiple blockchains on The Graph, requiring at least one subgraph of a layer-2 network for liquid restaking protocols.
\end{table}

We conducted the economic analysis with Renzo Protocol data for several reasons, instead of other LRPs including Ether.Fi. Renzo Protocol had an early and faster multi-blockchain adoption and LRT deployment on multiple blockchains, and it has more consistent and elaborative data availability (especially on The Graph). Since one of our aims is to assess the 'success' of a multi-blockchain business model (our 'success' proxy is the impact of the L2 TVL on the revenue in this case), it provides us a unique opportunity to observe this model in the early phase of this DeFi service. Additionally, we would like to analyze what is 'unexplored'. The liquid restaking market is currently dominated by Ether.fi, Renzo Protocol, and Kernel (Kelp) DAO. Ether.fi has been subjected to several studies, including \cite{b17} as the Ether.fi protocol had the highest TVL and immense growth. This market competition is involved in our model with a market share proxy. But analyzing competitor protocols in future research is still significant to generalize the findings.

\section{Research Questions and Hypotheses}

In this study, we first seek to answer the following questions. What are the primary determinants of revenue generation for a liquid restaking protocol, and how do multi-chain deployments impact revenue? We define the following hypotheses to test:
\begin{itemize}
    \item H1: The daily growth of Renzo Protocol's Ethereum TVL has a statistically significant and positive association with Renzo's revenue.
    \item H2: The TVL of the underlying EigenLayer infrastructure has a positive and statistically significant relationship with Renzo's protocol revenue.
    \item H3: The TVL of ezETH (Renzo's LRT) deployed on multiple blockchains has a positive and statistically significant relationship with Renzo's protocol revenue.
    \item H4: The yield rate of ezETH, as the yield of Renzo's liquid staking token, has a positive and statistically significant relationship with Renzo's protocol revenue.
    \item H5: The tokenization events (token generation announcements and events) have a structural and negative relationship with Renzo's revenue. 
\end{itemize}

After completing the statistical tests, we seek an answer to the following questions by an on-chain analysis. To what extent are the restaking service and LRTs interconnected to other DeFi services and blockchains? Do bridged and wrapped ezETHs pose a systemic risk to the ecosystem with current size?

\section{Methodology and Model}

Regression analysis is a common method to determine the relationship between dependent and independent variables in DeFi markets. Studies that have used regression analysis to assess various aspects in DeFi markets include ~\cite{b17}, ~\cite{b18}, and ~\cite{b19}. Ordinary least squares (OLS) regression estimates the linear relationship between variables by minimizing the sum of squared residuals between observed and predicted values. In this study, we first employ an OLS regression model to quantify the predictors of liquid-restaking revenue generation. Granger causality \cite{b20} tests and random forest feature importance \cite{b21} analysis follow the regressions to corroborate the findings independently. Granger-causality establishes temporal precedence. Random forest handles non-linearities and identifies true importance, and OLS models provide parametric estimates and significance testing. We believe that this methodological structure can provide us reinforced statistical results.

The baseline regression model is illustrated in Equation~\eqref{eq:baseline_model}. The baseline model is also tested with the lagged variables (\textit{T1} and \textit{T2}) to observe if there is a lagged effect. Hereby, the baseline model is called \textbf{Model 1}; the model with one-day lagged versions of all variables is called \textbf{Model 2}; and the model with two-day lagged versions of all variables is called \textbf{Model 3}. 

\begin{align}
\label{eq:baseline_model}
    Revenue_{t} &= \beta_0 + \beta_1 \mathrm{TVL0}_{t} + \beta_2 \mathrm{TVL1}_{t} + \beta_3 \mathrm{TVL2}_{t} \notag \\
       &\quad+ \beta_4 \mathrm{Yield}_{t} + \beta_5 \mathrm{Premium}_{t} + \beta_6  \mathrm{Share}_{t} \notag \\
       &\quad+ \beta_7 \mathrm{APY}_{t} + \beta_8 \mathrm{Events}_{t} + \beta_9 \mathrm{ETH}_{t} \notag \\
       &\quad   + \beta_{10} \mathrm{TxFee}_{t} + \beta_{11} \mathrm{FGI}_{t} + \epsilon_{t}
\end{align}

%\textbf{%
The dependent variable \textbf{(Revenue)} is the logarithm of the total revenue of Renzo Protocol in USD at time \( t \) on the Ethereum network.
%}% 
By 'revenue', we mean the total revenue of the protocol, involving the supply-side revenue. In additional tests made with protocol revenue and supply side revenue as dependent variables, we detected insignificantly small differences in the results since the protocol captures revenue at a fixed rate from the total revenue. For this reason, we use only the total revenue in our analysis. LRPs capture revenue on the Ethereum network even if the LRTs are deployed on multiple blockchains.

The model has several %\textbf{%
independent variables
%}% 
of interest. \textbf{\(\text{TVL0}_{t}\)} is the logged TVL in the EigenLayer protocol in USD at time \( t \) on the Ethereum network.
\textbf{\(\text{TVL1}_{t}\)} is the first log difference of TVL in the Renzo Protocol in USD at time \( t \) on the Ethereum network. \textbf{\(\text{TVL2}_{t}\)} is the aggregated and logged TVL in the Renzo Protocol in USD at time \( t \) on the layer-2 networks (Arbitrum, Base, Linea, Blast, and Mode). \textbf{\(\text{Yield}_{t}\)} is the yield rate of ezETH as it is the yield-bearing asset of Renzo's liquid restaking service at time \( t \). \textbf{\(\text{Premium}_{t}\)} is calculated as the percentage difference (deviation) of the ezETH price from the ETH price at time \( t \). \textbf{\(\text{Share}_{t}\)} is the market share of Renzo's LRT ezETH in the liquid restaking market at time \( t \), proxied by the total minted ezETH supply divided by the total amount of minted ETH in the LRPs.

The model has several %\textbf{%
control variables.
%}%
\textbf{\(\text{APY}_{t}\)} is the average APY rate of stETH as the benchmark yield-bearing asset in the DeFi markets at time \( t \). \textbf{\(\text{FGI}_{t}\)} is the daily change in the 'Fear And Greed Index' at time \( t \), used as a control variable to involve the general crypto market sentiment. \textbf{\(\text{ETH}_{t}\)} is the daily logarithmic return of the ETH price in USD at time \( t \). ETH is the benchmark asset used in the DeFi markets, specifically in staking services. \textbf{\(\text{TxFee}_{t}\)} is the volatility of the transaction fee log returns of the Ethereum network with 7-day rolling standard deviation at time \( t \). This variable is used for uncertainty or instability in the network. \textbf{\(\text{Events}_{t}\)} is the 
%\textbf{%
dummy variable
%}%
to observe the structural impact of tokenization events at time \( t \), involving the 'airdrop' tokenization announcements and token generation events of Renzo Protocol and EigenLayer network.

\section{Data and Feature Engineering}

The data of the EigenLayer and Renzo Protocols are collected from the subgraphs in The Graph ~\cite{b22}, a decentralized protocol for querying blockchain data. The data on ezETH yield and market share are collected from Dune Analytics ~\cite{b41}. The annual percentage yield (APY) data for stETH is collected from DeFiLlama ~\cite{b23}. The average transaction fee of the Ethereum network is collected from Etherscan ~\cite{b24}. ETH and ezETH price data is collected from CoinMarketCap ~\cite{b25}. The token supply balances of ezETH on the Linea network are collected from Lineascan ~\cite{b36}, not from The Graph since there is no subgraph for Renzo Protocol's Linea Network contracts. 'Crypto Fear \& Greed Index' data is collected from the API of Alternative ~\cite{b26}.

Tokenization events cover the days 26 April 2024 (Renzo Protocol tokenization announcement), 29 April 2024 (EigenLayer tokenization details are announced), 30 April 2024 (the REZ token was generated), and 1 October 2024 (the EIGEN token was generated).

The summary statistics describe raw economic magnitudes, not model reprocessing. The data, which are used in the economic analysis, cover the period between 22 January 2024 and 17 April 2025 with a daily time interval. The period starts from the data availability date for the Renzo Protocol on The Graph. Revenue, TVL values, ETH price and transaction fee variables are logarithmically transformed in regression analysis to mitigate skewness and interpret coefficients as elasticities, and help stationarity, as confirmed by the Dickey-Fuller Augmented Test (ADF)~\cite{b27}.

To overcome the methodological challenge (multiple-counting of TVL metrics), which is underlined by Chaudhry (2025)~\cite{b10} and Luo et al. (2025)~\cite{b12}, we used uninflated TVL data in our economic analysis. The TVL of the liquid restaking protocols is not involved in the TVL of EigenLayer to eliminate the common and inflated TVL since the LRT of the Renzo Protocol is also a restaked asset on the EigenLayer network. Similarly, the TVL of the Renzo Protocol does not involve the aggregated value of the bridged LRT on multiple blockchains, since the bridged ezETH (Renzo Protocol's LRT) is actually the representation of the ezETH on the Ethereum mainnet. Consequently, we have three TVL variables with uninflated numbers. We confirmed the suitable TVL data by comparing the data from several sources The Graph, DefiLlama, Etherscan, and the protocol websites.

\section{Results}

\begin{figure}[h]
    \centering
    \caption{\textbf{Regression Residuals Overtime:} The residuals of the baseline regression fluctuate randomly around zero with no apparent time-dependent structure, suggesting that the model adequately captures the mean behavior of the Renzo Protocol's $Revenue$ over time.}
    \includegraphics[width=0.5\textwidth]{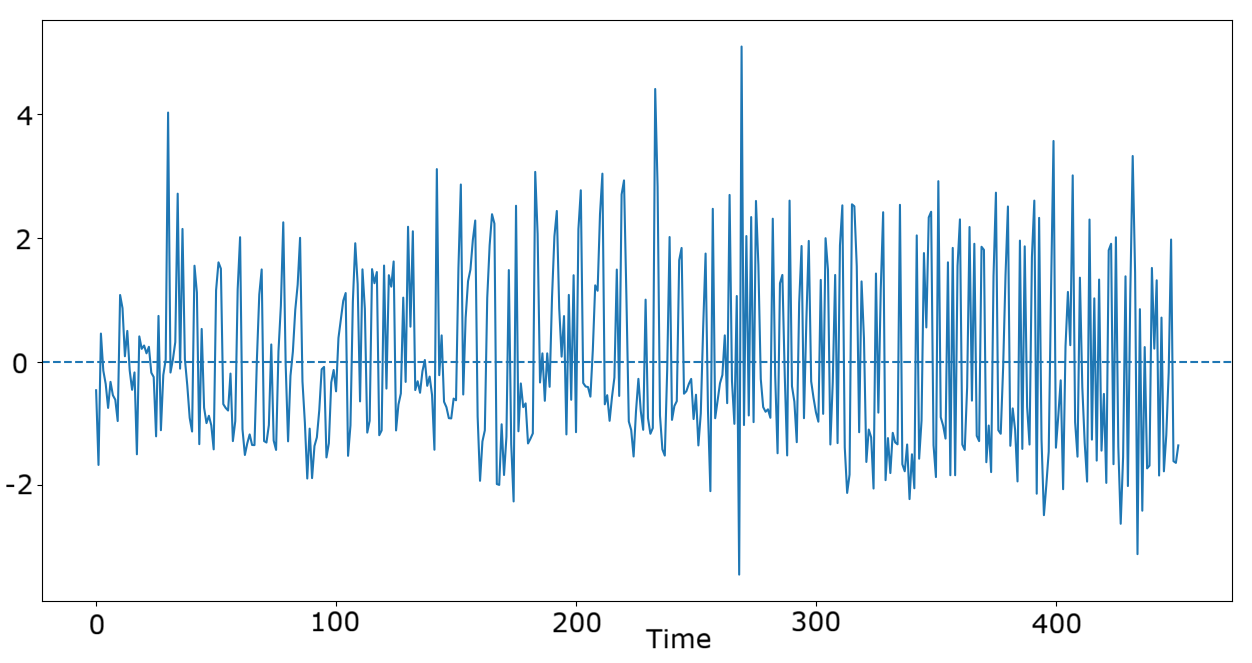}
    \label{fig:residuals_overtime_restaking.png}
\end{figure}

The summary statistics are introduced in Table \ref{tab:summary_statistics}. The variables do not have a mechanical correlation ($>$0.80) in the results of the correlation matrix. No high VIF value is observed, as introduced in Table \ref{tab:VIF}. The residuals are stationary and the regression residuals overtime is introduced in Fig. \ref{fig:residuals_overtime_restaking.png}. We took the difference of the initially non-stationary variables $TVL1$ and $ETH$. The OLS regression results are presented in Table \ref{tab:regression_results}. The results of the robustness test with winsorized data and robust standard errors (HC3) are presented in Table \ref{tab:robustness_results}. The results of Granger-causality tests are introduced in Table \ref{tab:regression_results}. The feature importance of variables is calculated using the random forest method presented in Table \ref{tab:feature_importance}.

\begin{table}
  \begin{threeparttable}
\centering
\caption{Summary Statistics}
\label{tab:summary_statistics}
\begin{tabularx}{\linewidth}{@{}Xccccccc@{}}
\toprule
\textbf{Metric} & \textbf{Mean} & \textbf{Std. Dev.} & \textbf{Min} & \textbf{Max} \\
\midrule
Revenue (\$)& 115.481 & 234.150 &	496 & 2.315.863 \\
TVL0 (\$mn)& 4.411 & 1.391 & 1.010 & 7.944 \\
TVL1 (\$mn)& 1.492 & 1.078 & 118	& 4.132 \\
TVL2 (\$mn)	& 169	& 216	& 0	& 756 \\
EigenLayer Users	& 564 &	2.367 &	5	& 41.125 \\
Renzo Ethereum Users & 139 &	297 &	1	& 2.026 \\
Renzo L2 Users &	2.939 &	3.399 &	0 &	55.284 \\
Yield	  & 3,509 &	6.557 &	-12,512 & 76,113 \\
Premium	& 0,0119 &	0,6936 &	-3,477	& 3.511 \\
Market Share	& 15,42 &	8,14 &	7,52 &	33,52 \\
APY	& 3,130 &	0,638 &	2,65 &	11,77 \\
ETH	(\$)& 2.947 &	589 &	1.472 &	4.065 \\
TxFee  & 4,680 &	4,751 &	0,33 &	29,5 \\
FGI	& 59,42 &	19,31 &	10	& 94 \\
\bottomrule
\end{tabularx}
\begin{tablenotes}
      \small
      \item \textit{Note:} The table introduces the summary statistics of the raw data with the 452 observations for the period of the regression analysis between 22nd January 2024 and 17th April 2025 with no missing days.
    \end{tablenotes}
  \end{threeparttable}
\end{table}

\begin{table}
  \begin{threeparttable}
  \centering
  \caption{OLS regression results}
  \label{tab:regression_results}
   \begin{tabular}{l@{\hspace{0.3cm}}r@{\hspace{0.4cm}}r@{\hspace{0.4cm}}r}
    \toprule
    & \textbf{Model 1} & \textbf{Model 2 (t-1)} & \textbf{Model 3 (t-2)} \\
    \midrule
    (Intercept)                       & -14.78** (4.55)       & -15.42*** ( 4.13)       & -11.94* (4.70) \\
    TVL0                            & 0.98*** (0.21)     & 1.00*** (0.19)     & 0.89*** (0.22) \\
    TVL1                            & -1.02 (1.86)     & -0.17 (1.68)     & -1.36 (1.90) \\
    TVL2                           & 0.08*** (0.02)     & 0.08*** (0.02)     & 0.07*** (0.02) \\
    Yield                   & 0.04*** (0.01)     & 0.10*** (0.01)     & 0.00 (0.01) \\
    Premium                 & 0.03 (0.10)     & -0.00 (0.09)     & -0.05 (0.11) \\
    Share                   & 0.07*** (0.01)     & 0.07*** (0.01)     & 0.06*** (0.01) \\
    APY                      & 0.21˙ (0.12)     & 0.23* (0.10)     & 0.09 (0.12) \\
    Events                 & -0.91 (0.76)     & -0.34 (0.69)     & -0.78 (0.78) \\
    ETH                      & 0.86 (1.97)     & 1.23 (1.78)      & -1.55 (2.01) \\
    TxFee                  & 0.42 (0.54)        & 0.23 (0.48)        & 0.02 (0.55) \\
    FGI                     & 0.02˙ (0.01)        & -0.02* (0.01)        & 0.02 (0.01) \\
    \midrule
    Observations                      & 452                & 451                & 450 \\
    R-squared                         & 0.314                & 0.438                & 0.275 \\
    \bottomrule
    \end{tabular}
\begin{tablenotes}
      \small
      \item \textit{Note:} This table presents the results of the OLS regressions with different lag structures. The dependent variable is \textit{\(Revenue_{t}\)}. The written values represent coefficients while statistical significance levels are shown with stars. Standard errors are reported in parentheses. Significance levels: ˙ p $<$ 0.10, * p $<$ 0.05, ** p $<$ 0.01, *** p $<$ 0.001.
    \end{tablenotes}
  \end{threeparttable}
\end{table}

\begin{table}
  \begin{threeparttable}
  \centering
  \caption{Robustness Check}
  \label{tab:robustness_results}
  \begin{tabular}{l@{\hspace{0.3cm}}r@{\hspace{0.4cm}}r@{\hspace{0.4cm}}r}
    \toprule
    & \textbf{Model 1} & \textbf{Model 2 (t-1)} & \textbf{Model 3 (t-2)} \\
    \midrule
    (Intercept)                       & -15.97** (5.70)       & -14.25** (5.29)       & -14.82* (6.24) \\
    TVL0                    & 0.97*** (0.26)     & 0.85*** (0.25)     & 1.00*** (0.29) \\
    TVL1    & -1.07 (2.22)     & -0.21 (2.00)     & -1.16 (2.27) \\
    TVL2               & 0.10*** (0.02)     & 0.094*** (0.01)     & 0.085*** (0.02) \\
    Yield                   & 0.11*** (0.02)     & 0.24*** (0.02)     & 0.03 (0.02) \\
    Premium                 & -0.10 (0.15)     & 0.07 (0.13)     & -0.15 (0.15) \\
    Share                   & 0.06*** (0.01)     & 0.07*** (0.01)     & 0.06*** (0.01) \\
    APY          & 0.50˙ (0.30)     & 0.69** (0.24)     & 0.11 (0.30) \\
    Events               & -0.70* (0.29)     & 0.01 (0.84)     & -0.71* (0.36) \\
    ETH          & 0.85 (2.51)     & 1.62 (2.00)      & -2.37 (2.57) \\
    TxFee               & 0.46 (0.65)        & 0.20 (0.55)        & -0.01 (0.68) \\
    FGI               & 0.03˙ (0.02)        & -0.02 (0.01)        & 0.01 (0.02) \\
    \midrule
    Observations                      & 452                & 451                & 450 \\
    R-squared                         & 0.339                & 0.519                & 0.273 \\
    \bottomrule
    \end{tabular}
\begin{tablenotes}
      \small
      \item \textit{Note:} This table presents the results of the robustness test with winsorized data and robust standard errors (HC3). $Events$ dummy variable is not winsorized. The written values represent coefficients. Standard errors are reported in parentheses. Significance levels: ˙ p $<$ 0.10, * p $<$ 0.05, ** p $<$ 0.01, *** p $<$ 0.001.
    \end{tablenotes}
  \end{threeparttable}
\end{table}

\begin{table}
\centering
\caption{Variance Inflation Factor (VIF) Results}
\label{tab:VIF}
\footnotesize
\begin{tabularx}{\linewidth}{
@{}lcc @{\hspace{1.5cm}} lcc@{}
}
\toprule
\textbf{Variable} & \textbf{VIF} & \textbf{Variable} & \textbf{VIF} & \textbf{Variable} & \textbf{VIF} \\
\midrule
TVL2   & 1.56 & TVL1 & 1.47 & TVL0   & 1.31  \\
 FGI   & 1.31 &  Share & 1.29 &  TxFee     &  1.16\\
 APY   & 1.12 & ETH & 1.05 & Events   & 1.05 \\
Premium   & 1.03 & Yield & 1.03 &       & \\
\bottomrule
\end{tabularx}
\begin{tablenotes}
\small
\item \textit{Note:} All VIF values are below 5, indicating no multicollinearity concerns.
\end{tablenotes}
\end{table}

\begin{table}
\centering
\caption{Granger Causality Test Results}
\label{tab:granger_causality_compact}
\footnotesize

\begin{tabularx}{\linewidth}{@{}lcc @{\hspace{1cm}} lcc@{}}
\toprule
\textbf{Variable} & \textbf{Lag} & \textbf{p-value} 
& \textbf{Variable} & \textbf{Lag} & \textbf{p-value} \\
\midrule

TVL2 & 1 & 0.0000***   & TxFee  & 2 & 0.1436 \\
TVL0 & 1 & 0.0000***   & APY  & 3 & 0.1496 \\
Share & 1 & 0.0000***   & ETH  & 5 & 0.1550 \\
Yield & 4 & 0.0000***   & Premium & 5 & 0.2393 \\
FGI & 1 & 0.0048**     & Events  & 4 & 0.2778 \\
TVL1 & 1 & 0.0419*      &    &   &      \\

\bottomrule
\end{tabularx}

\begin{tablenotes}
\small
\item \textit{Note:} The Granger causality results are introduced in the table. $TVL0$, $TVL2$, $Share$ and $Yield$ show statistically significant predictive content in their relationships with $Revenue$. Significance levels: *** p $<$ 0.01, ** p $<$ 0.05, * p $<$ 0.10.
\end{tablenotes}

\end{table}

\begin{table}
  \begin{threeparttable}
\centering
\caption{Feature Importance Results}
\label{tab:feature_importance}
\footnotesize

\begin{tabularx}{\linewidth}{@{}lcc @{\hspace{0.5cm}} lcc@{}}
\toprule
\textbf{Feature} & \textbf{Gini} & \textbf{Permutation}
& \textbf{Feature} & \textbf{Gini} & \textbf{Permutation} \\
\midrule
TVL2   & 0.2589 & 0.6165   & TxFee  & 0.0732 & 0.0845 \\
Yield   & 0.1406 & 0.3771   & TVL1  & 0.0635 & 0.0801  \\
TVL0  & 0.1228 & 0.1834 & ETH   & 0.0664 & 0.0791 \\
Share   & 0.0871 & 0.1336  & FGI   & 0.0535 & 0.0738 \\
APY   & 0.0713 & 0.0959  & Premium   & 0.0626 & 0.0703 \\
\bottomrule
\end{tabularx}

\begin{tablenotes}
\small
\item \textit{Note:} The results of random forest feature importance are presented in the table. The most important feature is $TVL2$ to analyze the drivers of $Revenue$. $Yield$ and $TVL0$ follows it. The random forest method prioritizes continuous and high-variance predictors. This is the reason the $Events$ dummy variable is not included.
\end{tablenotes}

  \end{threeparttable}
\end{table}

The results reveal a nuanced picture, confirming several foundational hypotheses and contradicting some others. The analysis provides strong evidence supporting the three core hypotheses regarding ecosystem growth.

The results support H2. We find that EigenLayer's TVL has a statistically significant association with LRP revenue according to the results (OLS p$<$0.001, Granger p=0.0000, Random Forest Importance: 0.1228). $TVL0$ exhibits a positive relationship with the revenue (1\% increase in EigenLayer's TVL is associated with a 0.98\% increase in the revenue for the baseline model). This finding may indicate that the growth of the entire restaking ecosystem acts as a rising tide that lifts all boats.

H3, which posited a positive and significant relationship between Renzo's aggregated L2 TVL and revenue, is supported. The significance of $TVL2$ in all applications confirms that L2 expansion is a significant predictor of revenue dynamics (OLS p$<$0.001, Granger p=0.0000, Random Forest Importance: 0.2589). 1\% increase in Renzo's L2 TVL is associated with a 0.08\% increase in the revenue for the baseline model. The results may show an alignment with the theoretical view that multi-blockchain expansion broadens the protocol's reach and revenue base.

The results show that there is no reliable relationship between Renzo's main Ethereum network TVL and the revenue (OLS insignificant, Granger p=0.0419, Random Forest Importance: 0.0635). The consistent lack of a statistically significant relationship between $TVL1$ and $Revenue$ is an unexpected finding. H1 is firmly rejected. We observe a low active user activity on Renzo's Ethereum protocol, while a high active user activity of Renzo Protocol on L2 networks is seen in Table \ref{tab:summary_statistics}. This user activity difference might explain the divergent relationship of Renzo Protocol's TVL variables with the revenue. Together, these results may underscore the paramount importance of user and network adoption.

H4 is supported by empirical evidence. The results show a positive and significant relationship between the ezETH $Yield$ rate and Renzo's revenue (OLS p$<$0.001, Granger p=0.0000, Random Forest Importance: 0.1406). A one-unit (one percentage-point) increase in $Yield$ is associated with approximately 4\% increase in revenue for the baseline model. There is significant change in the yield check of the robust Model 2: a one-unit increase in $Yield$ is associated with a 27.14\% increase in revenue. Thus, the results may indicate that Renzo's revenue is tethered to its LRT's yield rate.

The market $Share$ variable, which captures Renzo's competitive position relative to Ether.fi and other LST protocols, serves as an indirect cross-protocol control and is statistically significant (OLS p$<$0.001, Granger p=0.0000, Random Forest Importance: 0.0871). A one percentage-point increase in $Share$ is associated with a 7.25\% increase in revenue for the baseline model. The weak relationship between $APY$ as the benchmark yield of DeFi (stETH APY) and $Revenue$ may still provide a financial anchor connecting liquid restaking to the established staking market. Meanwhile, the lack of significance for broader market controls like ETH price and transaction fees suggests that, Renzo's revenue is more insulated from general crypto market fluctuations and is primarily driven by its service dynamics and the specific growth trajectory of the restaking ecosystem.

H5, which posits the dummy variable of tokenization events as statistically significant in the model, is rejected since the applied tests do not indicate a significant result for $Events$ variable in the baseline model. However, the Chow test indicates a structural break for the EIGEN token generation event on October 1, 2024.

The explanatory power and R-squared of the regression models peak in the robustness test of Model 2 with winsorized data and robust standard errors (HC3).

Lastly, in the regression analysis, no significant relationship is detected between $Premium$, as an LRT price premium proxy, and $Revenue$.

\section{Connected Risk of Restaking and Bridge}

The emergence of liquid restaking introduces a new layer of interconnected risk to the liquid-staking ecosystem and the cryptoeconomic security of Ethereum. At its core, we are referring to a risk to the Ethereum main network's security, which is provided by the crypto-economic staking mechanism. Leveraging the liquid usage of the staked ETHs across the layered DeFi services also pushes the impact of a loss in the crypto-economic value beyond the fundamental value of the natively staked assets.

The threat, as highlighted by Ethereum co-founder Vitalik Buterin, is a cascading slashing \cite{b28}. This is not just a protocol-level failure; it is a systemic one. Quantifying this risk is challenging, but observing the scale is critical to enable scenario-based stress testing -such as estimating token depegs, liquidity shortfalls, or the diffusion of LRTs to the staking under hypothetical shocks- and quantifies the economic impact of these events on the Ethereum network security. The approach informs protocol design decisions, such as deciding collateralization ratios, introducing withdrawal buffers, or adjusting staking incentives to reduce systemic vulnerability.

\begin{figure}[h]
    \centering
    \caption{\textbf{Value Flow Chart:} represents the interconnectedness between different DeFi services, specifically staking and restaking services. The numbers represent the ETH equivalent amounts as of 4th October 2025 12:00 PM UTC. The figure does not cover the overall decentralized finance ecosystem. However, it covers the top DeFi services and the related protocols that this paper focuses on, to underline the layered risks of bridges, rollups, and other protocols. Sources: \href{https://beaconcha.in/charts/staked_ether}{Beaconcha.in}, \href{https://etherscan.io/token/0xae7ab96520de3a18e5e111b5eaab095312d7fe84}{Etherscan}, \href{https://app.aave.com/reserve-overview/?underlyingAsset=0x7f39c581f595b53c5cb19bd0b3f8da6c935e2ca0&marketName=proto_mainnet_v3}{AAVE}, \href{https://app.uniswap.org/explore/pools/ethereum/0x109830a1aaad605bbf02a9dfa7b0b92ec2fb7daa}{Uniswap}, \href{https://www.curve.finance/dex/ethereum/pools/factory-tricrypto-2/deposit}{Curve Finance}, \href{https://balancer.fi/pools/ethereum/v2/0x3de27efa2f1aa663ae5d458857e731c129069f29000200000000000000000588}{Balancer}, \href{https://app.pendle.finance/trade/pools/0xc374f7ec85f8c7de3207a10bb1978ba104bda3b2/zap/in?chain=ethereum}{Pendle Finance}, \href{https://defillama.com/protocol/renzo?denomination=ETH}{DeFiLlama}, \href{https://app.aave.com/reserve-overview/?underlyingAsset=0x2416092f143378750bb29b79ed961ab195cceea5&marketName=proto_linea_v3}{AAVE v3 Linea}, \href{https://lineascan.build/address/0xd66d0e2454d9e0eee51440cd23215f46e7d20a83}{Etherex} (Last Access: October 4, 2025).}
    \vspace{0.5cm}
    \includegraphics[width=0.48\textwidth]{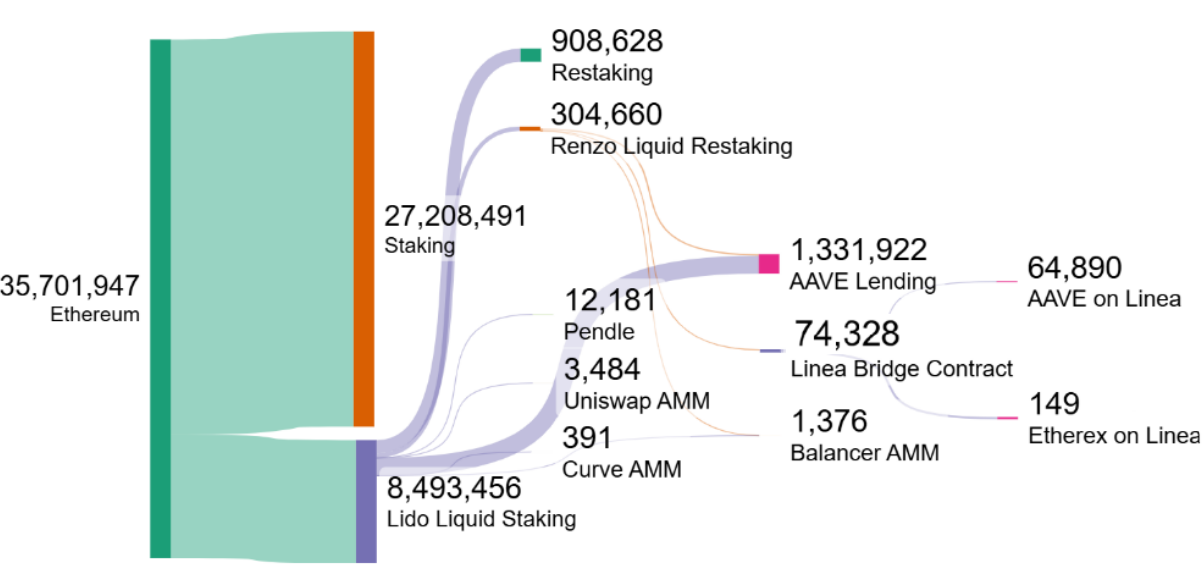}
    \vspace{-0.5cm}
    \label{fig:value_flow_chart}
\end{figure}

The interconnected value flows and risks are shown in Fig. \ref{fig:value_flow_chart}. The total restaking market involves 19.23B USD worth of TVL (natively restaked asset TVL is 13.97B USD while liquid-restaked assets have 4.06B USD TVL) \cite{b23}. This total restaked TVL has approximately 13\% of the underlying staked ETH, since the number of staked ETH is 35,701,947 as of October 9, 2025 \cite{b30}. Since the percentage of restaking is low for the staked ETHs in comparison to the total amount of staked ETH, we cannot claim that a cascading slashing through restaking would create a network security risk for Ethereum as the chain requires over 33\% of the staked ETH to sustain the current crypto-economic security (preventing attacks on finality), as of 9 October 2025 \cite{b31}. The overview of the staked and restaked ETH numbers can be seen in Table \ref{fig:value_flow_chart}.

On the other hand, we observe that most of the restaked assets in EigenLayer are natively staked ETH with 72.79\% of all restaked assets with value as of October 9, 2025, rather than stETH, which is the widely used LST across different DeFi services \cite{b32}. Thus, one can claim that the security risks of the liquid-staking protocols do not threaten the restaking service and the crypto-economic security of the network lethally.

LRT such as ezETH on layer-2 networks and BSC are bridged or wrapped assets \cite{b33}. The bridging process typically involves using a bridge contract to lock the native LRT on its home chain and mint a representative version on the destination chain. These bridge contracts undoubtedly pose classical bridge risks to the restaking ecosystem. Since every restaked ETH has an underlying staked ETH, a bridge failure could potentially cause a future slashing cascade.

A severe slashing event that significantly devalues major LSTs could trigger a massive sell-off. As the value of the collateral backing DeFi loans plummets, it could spark a systemic liquidity crisis throughout the ecosystem. This does not make it easier to 'attack the blockchain' in a traditional 51\% sense, but it could destabilize the economic layer that justifies and secures the network, since LSTs and LRTs are used in different DeFi protocols, as seen in Fig. \ref{fig:value_flow_chart} cross-protocol contagion via DeFi exposure. LSTs are widely used in lending pools, automated market makers (AMM) pools, and yield-optimization protocols.

\textbf{Bridge exploits} are among the most devastating hacks, as bridge contracts are high-value targets \cite{b23}. We argue that bridged LRTs create a dangerous cross-effect: a bridge exploit could drain a significant portion of LRTs, destabilizing their peg and causing panic, while simultaneously, a slashing event on the main chain could render the bridged tokens on L2s worthless. The total amount of ezETH on Ethereum is 304.660. According to the amount of ezETH on the external blockchain network on 9 October 2025, a total of 90.862,04 ezETH is monitored on the Linea, Arbitrum, Base, Blast and BSC blockchains \cite{b23}. This amount is approximately 29.82\% of the total ezETH of the Renzo Protocol, approximately 2.1\% of the total LRT market cap and 2\% of the total restaked ETH market cap. As of 9 October 2025, the size and amount of ezETH currently do not impose a significant risk to the overall restaking network and the crypto-economic security of the Ethereum network.

If a single rollup becomes the dominant home for LRTs, the rollup risks shifting to restaking protocols and risks, such as \textbf{sequencer centralization risk} (which also introduces the possibility of transaction censoring), proof-validation risk, and compromised multi-sig contracts. At this point, the decentralization status of a bridged rollup is critical. Linea is the layer-2 network that recently has the highest amount of ezETH (74.328,67 ezETH, which is approximately 81.80\% of the total bridged ezETH) among the bridged blockchains. Since the ezETH bridged to Linea includes approximately 24\% of total ezETH, the rollup risks of Linea can create a direct impact on the Renzo Protocol \cite{b36}. However, the Renzo Protocol would not have a direct impact on the overall restaking and staking ecosystem with respect to its current size, as explained previously.

If one liquid-restaking protocol dominates the market, it may pose both \textbf{restaker and node operator centralization risks}. A single protocol acting as the primary issuer of LRTs would concentrate the power to delegate vast amounts of staked ETH. This concentration may create a single point of failure; misbehavior or compromise of a major operator or protocol could trigger the cascading slashing events previously described, impacting a significant downfall of the ecosystem simultaneously. As of 9 October 2025, Ether.fi's eETH is the dominant liquid-restaking protocol with 65.4\% of total LRTs and it represents 62.81\% of the total restaked value \cite{b24}. The design of leading protocols such as Etherfi, Puffer, and Renzo aims to mitigate the risks above, but market forces could naturally lead to consolidation, undermining decentralization in terms of protocol and asset diversity.

\section{Scenarios of Contagious Interconnectedness}

Considering and quantifying scenarios that overload Ethereum's crypto-economic security becomes more important \cite{b28} to design network and ecosystem economics for the new complexities created by a layered and interconnected DeFi and multi-blockchain architecture.

A contagious situation can be considered with a hypothetical scenario of compromised LRTs. In another scenario, a logic failure in the smart contract of an AMM's liquidity pool on the bridged network results in the irreversible loss or mispricing of ezETH deposited by liquidity providers. If the failure occurs on the bridged network, recovery is further complicated by cross-chain settlement delays and liquidity fragmentation.

To quantify contagion risk, we conduct a liquidation stress test using on-chain data and Aave v3 risk parameters for ezETH on Linea ($LTV = 72.5\%$; $LT = 75\%$). Although ezETH borrowing is currently disabled with $0\%$ reserve utilization, we assume borrowing is enabled at the maximum $LTV$ (worst-case), so the debt equals $\text{Debt} = LTV \times \text{Collateral}$. We further assume WETH or USDC can be borrowed against ezETH collateral, with the Chainlink oracle tracking ezETH/ETH. Under a fractional price decline $\delta$, the debt is unchanged and the health factor becomes:
\begin{equation}
HF(\delta) = (1 - \delta)\cdot\frac{LT}{LTV}
\label{eq:hf_depeg}
\end{equation}
Liquidation triggers when $HF < 1$. Setting $HF = 1$ yields the critical depeg threshold:
\[
\delta^{*} = 1 - \frac{LTV}{LT} = 1 - \frac{0.725}{0.75} \approx \mathbf{3.33\%}.
\]
Thus, a decline of just $3.33\%$ in bridged ezETH renders all maximum-LTV positions liquidatable. As of 4 October 2025 12:00 PM UTC, $64{,}890$~ezETH is posted as collateral in the Aave V3 Linea pool. A depeg at or beyond this threshold would expose the entire pool to liquidation, forcing up to $64{,}890$~ezETH onto the market. Local DEX liquidity on Linea (Etherex) holds only $149$~ezETH, a coverage ratio of $149 / 64{,}890 \approx 0.23\%$, so sell pressure would bridge back to Ethereum mainnet. There, the combined ezETH and stETH liquidity across Uniswap, Balancer, and Curve amounts to $\approx 5{,}251$~ETH equivalent — covering only $5{,}251 / 64{,}890 \approx 8.09\%$ of the potentially liquidated collateral, indicating substantial systemic price impact under either attack scenario (smart contract failure or bridge compromise). If liquidations propagate to Lido (which holds $8{,}493{,}457$ staked ETH; see \ref{fig:value_flow_chart}) via redemption pressure on restaked assets, the forced unwind corresponds to $64{,}890 / 8{,}493{,}457 \approx 0.76\%$ of liquid-staked ETH in the ecosystem. While unlikely to directly threaten Ethereum's proof-of-stake consensus, this could generate short-term validator exit pressure and amplify volatility through cascading liquid staking and restaking redemptions.

This cascading decline reduces the collateral value of ezETH in lending markets and triggers potential liquidations, while liquidity pools experience imbalance and increased slippage. Bridged deployments worsen these effects due to fragmented liquidity and settlement frictions across networks. The hack of Balancer protocol, which resulted in resulted 120M USD worth of crypto-asset loss, has clearly shown that high value of cross-contamination is always on the table for tokenized, layered and interconnected DeFi services \cite{b40}.

\section{Limitations}

This study acknowledges several important limitations. First, the moderate explanatory power of our models indicates significant unexplained variance. Second, the analysis captures a specific market phase in the rapidly evolving restaking landscape, potentially limiting the generalizability of findings across different market cycles. We specifically focus on one protocol in our analysis, as the business model, protocol positioning, and the available data of the Renzo Protocol provide us with a unique opportunity to investigate several perspectives in a study. The relationships identified in the regression analysis are exploratory and do not imply causation. Co-integration tests might be required to understand the relationships better. The observed dynamics may not apply to protocols with different structures or operational models. Fourth, the analysis relies on available subgraph data, which may include inconsistent values. Additionally, restaking service is considerably new, which means historical data on protocols is limited, potentially constraining the analysis of long-term trends in stability. The fast-paced development of the DeFi space may quickly outdate the findings of this study.

\section{Conclusion}

Our analysis of revenue dynamics revealed a critical insight: not all growth is equal. A powerful predictor of the Renzo Protocol's revenue is the value in the underlying EigenLayer protocol, not the value locked in the liquid-restaking protocol. Furthermore, expanding to multiple blockchain networks by deploying the liquid-restaking token on many blockchains has statistically significant and positive relationship with revenue. The yield rate of the liquid-restaking token has a significant and strong association with the protocol revenue. Surprisingly, the tokenization events like 'airdrop' announcements and token generations does not show a significant relationship with revenue. The overall economic analysis shows that the revenue model is driven by multi-blockchain deployment, ecosystem growth, and the yield rate, rather than a simple speculation of the liquid restaking token. The study shows that the economic model of a DeFi protocol can bring deeply connected security complexities. From a technical point of view, the pursuit of higher yields through liquid restaking comes with a clear trade-off. The same strategy that boosts revenue also spreads and compounds risk across the DeFi ecosystem and multi-blockchain environment. This very expansion spreads several risks simultaneously, including bridge risk and the interconnectedness of different DeFi services and blockchains. Although the current size of the Renzo protocol and the overall restaking ecosystem is not large enough to threaten Ethereum's core security, it creates a chain of dependency. In the future, the compounded overload of leveraged services on Ethereum's crypto-economic security may impose bigger risks on the network security. A sustainable economic growth requires a balanced approach with a clear understanding and management of the interconnected risks.


\end{document}